\documentclass[%
 reprint,
 superscriptaddress,
 amsmath,
 amssymb,
 prl,
]{revtex4-1}

\usepackage{graphicx}
\usepackage{dcolumn}
\usepackage{bm}
\usepackage{color}
\usepackage{subcaption}

\begin{document}

\title{Stabilisation of the Arrival Time of a Relativistic Electron Beam to 
the 50~fs Level}

\author{J.~Roberts}
\affiliation{John Adams Institute (JAI), University of Oxford, Denys Wilkinson 
Building, Keble Road, Oxford, OX1 3RH, United Kingdom}
\affiliation{The European Organization for Nuclear Research (CERN), Geneva 23, 
CH-1211, Switzerland}

\author{P.~Skowronski}
\affiliation{The European Organization for Nuclear Research (CERN), Geneva 23, 
	CH-1211, Switzerland}

\author{P.N.~Burrows}
\affiliation{John Adams Institute (JAI), University of Oxford, Denys Wilkinson 
Building, Keble Road, Oxford, OX1 3RH, United Kingdom}

\author{G.B.~Christian}
\affiliation{John Adams Institute (JAI), University of Oxford, Denys Wilkinson 
Building, Keble Road, Oxford, OX1 3RH, United Kingdom}

\author{R.~Corsini}
\affiliation{The European Organization for Nuclear Research (CERN), Geneva 23, 
CH-1211, Switzerland}

\author{A.~Ghigo}
\affiliation{Laboratori Nazionali di Frascati (LNFN), Via Enrico Fermi, 40, 
00044 Frascati RM, Italy}

\author{F.~Marcellini}
\affiliation{Laboratori Nazionali di Frascati (LNFN), Via Enrico Fermi, 40, 
00044 Frascati RM, Italy}

\author{C.~Perry}
\affiliation{John Adams Institute (JAI), University of Oxford, Denys Wilkinson 
Building, Keble Road, Oxford, OX1 3RH, United Kingdom}

\date{\today}

\begin{abstract}
We report the results of a low-latency beam phase feed-forward system built to 
stabilise the arrival time of a relativistic electron beam. The system was 
operated at the Compact Linear Collider (CLIC) Test Facility (CTF3) at CERN 
where the beam arrival time was stabilised to approximately 50~fs. The 
system 
latency was \(350\)~ns and the correction bandwidth \(>23\)~MHz. 
The system meets the requirements for CLIC.
\end{abstract}

\maketitle

High-energy linear electron-positron colliders have been proposed as 
next-generation particle accelerators for exploring the subatomic world with 
increased precision. They provide sensitivity to new physics processes, 
beyond those described by the Standard Model (SM) of elementary particle 
interactions, at mass scales that can exceed the reach of the CERN 
Large Hadron Collider (LHC) \cite{CLIC-staging}.

The Compact Linear Collider (CLIC)~\cite{CLICCDR} is the most technologically 
mature concept of a high-energy lepton collider for enabling direct searches 
for new physics in the multi-TeV energy regime. It uses a novel two 
beam acceleration concept to achieve a high accelerating gradient of 100 MV/m 
and centre-of-mass collision energies of up to 3~TeV. This energy reach, 
combined with high-luminosity of the electron-positron collisions, will also 
enable precise measurements of properties of the Higgs boson~\cite{CLIC-Higgs} 
and the top quark, and provide sensitivity to beyond-SM 
phenomena~\cite{CLIC-staging}.

\begin{figure}
	\includegraphics[width=\columnwidth]{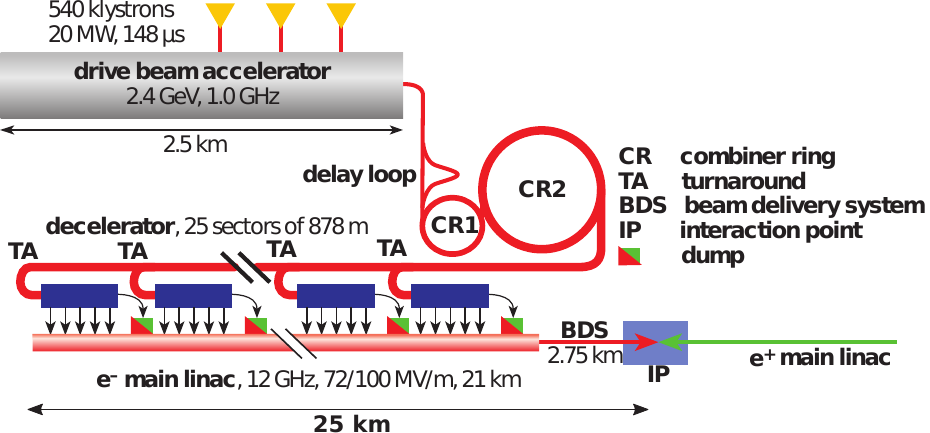}
	\caption{\label{fig:CLICLayout} Schematic of the CLIC drive-beam 
	concept showing the electron acceleration complex \cite{CLIC-staging}.
	}
\end{figure}

The CLIC two-beam acceleration concept is shown schematically in 
Fig.~\ref{fig:CLICLayout}. 
The 12~GHz RF power used to accelerate the colliding electron and positron 
beams is extracted from high intensity `drive beams'. 
The drive beams are 2.4~GeV electron beams, with an initial bunch frequency of 
0.5~GHz, a pulse length of \(148~\mathrm{\mu s}\), and a pulse repetition rate 
of 50~Hz. The intensity of the drive beams is increased by a factor 24 using a 
bunch recombination process \cite{CLICCDR}, thereby creating a series of 240~ns 
pulses bunched at 12~GHz.
Each 240~ns sub-pulse is directed into a `decelerator sector', in which the 
drive beam pulse is decelerated, producing 12~GHz RF power which is transferred 
to the accelerating structures of the main beams. Two drive beams with 25 
decelerator sectors each are required for a 3~TeV collider.

One of the major challenges is the synchronisation of the arrival of the drive 
and main beams at the power-extraction and transfer structures to better than 
50~fs rms. This requirement limits the luminosity loss, resulting from 
subsequent energy errors of the main beams, to less than 1\% of the design 
value~\cite{clicLumEq}. Free-electron lasers (FELs) also demand a high degree 
of beam arrival-time stability w.r.t. an externally-applied laser beam for the 
purpose of seeding of lasing by the electron beam \cite{Savelyev2017}. 

We express the temporal stability of the drive beam in terms of phase 
stability at the 12~GHz acceleration frequency. An arrival time jitter of 50~fs 
rms is equivalent to a phase jitter of \(0.2^\circ\)~at~12~GHz.
In the CLIC design the incoming drive-beam phase jitter 
cannot be guaranteed to be better than \(2^\circ\)~\cite{CLICCDR}. A mechanism 
to improve the phase stability by an order of magnitude is 
therefore required. The correction must be applied to the full drive beam pulse 
length and have a bandwidth exceeding 17.5~MHz. This bandwidth is derived from 
simulations of the system performance whilst assuming a pessimistic frequency 
spectrum of the incoming phase errors~\cite{Gerber2015}. 

This is implemented via a `phase feed-forward' (PFF) system which measures the 
incoming beam phase and provides a correction to the same beam pulse 
after it has traversed the turnaround loop (TA in Fig.~\ref{fig:CLICLayout}). 
One PFF system will be installed in each deceleration sector. The correction 
is provided by electromagnetic kickers in a 4-bend chicane: bunches arriving 
early (late) in time have their path through the chicane lengthened (shortened) 
respectively. A particular challenge is that the PFF latency must be shorter 
than the beam flight time of approximately.~250~ns around the turnaround loop.

\begin{figure}
	\includegraphics[width=\columnwidth]{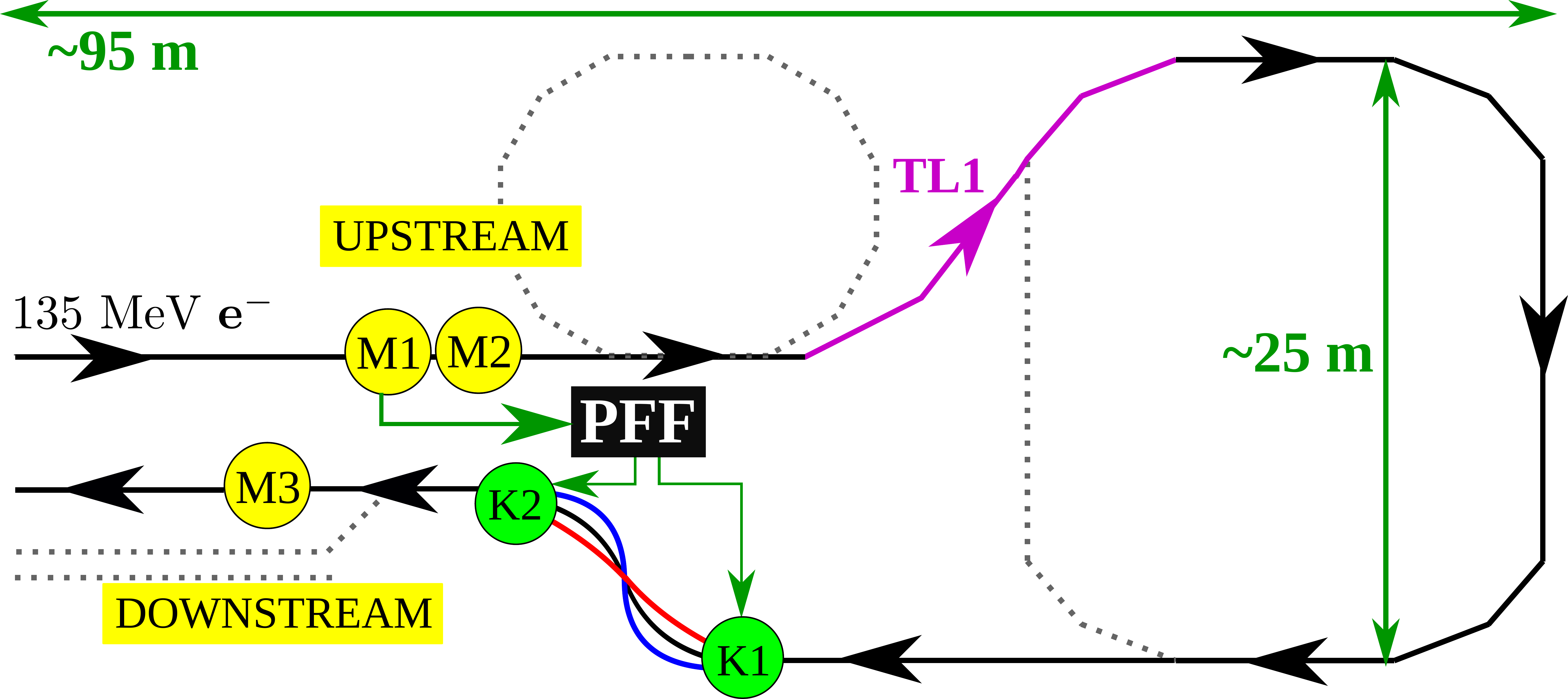}
	\caption{\label{fig:pffLayout}Schematic of the CTF3 PFF prototype, 
	showing the phase monitors (\(M_1\) , 
	\(M_2\) and \(M_3\)) and kickers (K1 and K2). The black box “PFF” 
	represents the calculation and output of the correction. Bunches arriving 
	early at \(M_1\) are deflected on to longer trajectories in the 
	chicane (blue), and bunches arriving late on to shorter trajectories (red). 
	Dashed lines indicate beam lines that are not used. 
		}
\end{figure}

We describe a prototype PFF system (Fig.~\ref{fig:pffLayout}) that implements 
this novel concept at the CLIC Test Facility (CTF3) at CERN. CTF3 provides a 
135~MeV electron beam bunched at 3~GHz frequency with a beam-pulse length of 
1.2~\(\mathrm{\mu s}\) and a pulse repetition rate of 0.8~Hz \cite{CLICCDR}. 

The incoming beam phase is measured in two upstream phase 
monitors (\(M_{1}, M_{2}\)). While the beam 
transits the ‘turnaround loop’ a phase-correction signal is evaluated and used 
to drive fast, high power amplifiers; these drive two electromagnetic kickers 
(\(\mathrm{K_1, K_2}\)) which are used to alter the beam transit time in a 
four-bend chicane. A downstream phase monitor (\(M_{3}\)) is 
used to measure the effect of the correction. 

The beam time of flight between \(M_1\) and \(\mathrm{K_1}\) is around 
380~ns. The total cable delay for the PFF correction signals 
is shorter, around 250~ns. The correction in the chicane can therefore be 
applied to the entirety of the beam pulse measured at the PFF input 
(\(\phi_1\), the measured phase at \(M_1\)), provided that the hardware
latency is less than 130~ns. 
Significant hardware challenges include the resolution and bandwidth of the 
phase monitors, and the power, latency and bandwidth of the kicker amplifiers. 
A low latency digitiser/feedforward controller is also required.
 
The requirements of the CLIC system and their corresponding CTF3 values are 
listed in Table~\ref{tab:pffspecs}. The main differences result from the 
different drive-beam energies. Higher power 
amplifiers (500~kW rather than 20~kW) are required for CLIC, which may be 
achieved by combining the output of multiple modules similar to those built for 
CTF3.
CLIC also requires a distributed timing system to synchronise the phase of the 
drive and main beams along the 50~km facility, which is not addressed here.

\begin{table}
	\caption{\label{tab:pffspecs}
	    Requirements for the CLIC PFF system, and the respective CTF3 
	    parameters; performance achieved with the prototype system is indicated 
	    by \textbf{*}.}
\begin{ruledtabular}
	\begin{tabular}{lccc}
		 & CLIC & CTF3 \\
		\hline
		Drive beam energy & 2400 & 135 & MeV \\
		No. PFF systems & 50 & 1 & \\
		Kickers per PFF chicane & 16 & 2 & \\
		Power of kicker amplifiers & 500 & \(\mathbf{20^*}\) & kW \\
		Angular deflection per kicker & \(\pm94\) & 
		\(\mathbf{\pm560^*}\) & \(~\mathrm{\mu rad}\) \\
		Correction range & \(\pm 10\) & \(\mathbf{\pm 6^*}\) & \(^\circ\) \\
		Correction bandwidth & \(>17.5\) & \(\mathbf{>23^*}\) & MHz \\
		Phase monitor resolution & \(< 0.14\) & \(\mathbf{0.12^*}\) &  
		\(^\circ\)   \\
		Initial phase jitter & \(2.0\) & \(0.9\) &  \(^\circ\) \\
		Corrected phase jitter & \(0.2\) & \(\mathbf{0.2^*}\) &  \(^\circ\)  \\
	\end{tabular}
\end{ruledtabular}
\end{table}

The phase monitors~\cite{phMonEuCard} are cylindrical cavities with an aperture 
of 23~mm and a length of 19~cm. Small ridges (’notch filters’) in the cavity 
create an effective volume with a resonant frequency of 12~GHz. 
The field induced by the beam traversing the cavity 
contains a beam-position-independent monopole mode and an unwanted 
position-dependent dipole mode. The 
effect of the latter is removed by summing the outputs from an opposing pair 
of feedthroughs, on the top and bottom of the cavity, via a RF ‘hybrid’. 
To extract the beam phase the output from each hybrid 
is mixed with a 12~GHz reference signal derived from a 3~GHz source which is 
phase-locked to the CTF3 RF system and serves all three phase monitors.
By comparing the signals from \(M_1\)~and~\(M_2\) we have measured a 
phase resolution of \(0.12^\circ\), i.e. about 30~fs~\cite{RobertsThesis}.

The phase signals are digitised  in the feedforward controller
board~\cite{RobertsThesis}, which is used to calculate and output the amplifier 
drive signals, and to  control the correction timing. It consists of nine 
14-bit analogue to digital converters clocked at 357~MHz, a field programmable 
gate array, and four digital to analogue converters. 

The kicker amplifiers~\cite{RobertsThesis} consist of one central control 
module and two drive and terminator modules (one per kicker). The control 
module distributes power and input signals to the 
drive modules. The 20~kW drive modules consist of low-voltage Si FETs driving 
high-voltage SiC FETs; an input voltage range of \(\pm2\)~V corresponds to an 
output range of \(\pm700\)~V. The response is linear to within 3\% for input 
voltages between \(\pm1.2\)~V, and the output bandwidth is 47~MHz for small 
signal variations of up to 20\% of the maximum. For larger signal variations 
the bandwidth is slew-rate limited.

The two electromagnetic stripline kickers \cite{kickerIPAC11} are 1~m in length 
and have an internal aperture of 40~mm between two strips placed along their 
horizontal walls. They are designed to give a response within a few ns of the 
input signal.
Opposite polarity voltages of up to 700~V applied to the 
strips at the downstream end horizontally deflect the 135~MeV 
beam by up to 560~\(\mu\)rad.

The measured total latency of the phase monitor signal processing, the 
feedforward calculation, and amplifier response was approximately 100~ns. 
Therefore the output from the controller was delayed by an additional 30~ns to 
synchronise the correction at the kicker with the beam arrival 
\cite{RobertsThesis}.

The PFF operation placed severe constraints on the setting of the 
magnetic lattice in both the beamline between the upstream phase monitors and 
the correction chicane, and in  the chicane itself.
The beam transfer matrix coefficient \(R_{52}\) between the two kickers 
characterises the change in path length through the chicane relative to the 
deflection applied at the first kicker. 
With an \(R_{52}\) value of \(0.74\)~m/rad \cite{RobertsThesis} the expected 
maximum path length change for operation of the PFF system, corresponding to 
the maximum deflection of \(\pm560\)~\(\mu\)rad from each kicker, is about 
\(\pm400~\mathrm{\mu m}\), equivalent to \(\pm6^\circ\) in phase. 
The chicane magnets were also set so that PFF operation does not change the 
beam trajectory at the exit of the chicane~\cite{RobertsThesis}.

A further challenge to PFF operation was obtaining a high correlation 
between the upstream and uncorrected downstream phases measured at \(M_1\) 
and \(M_3\) respectively. 
The maximum measurable correlation depends on both the phase monitor resolution 
and any additional phase jitter introduced in the beamlines between \(M_1\) 
and \(M_3\). The monitor resolution of \(0.12^\circ\) limits the maximum 
upstream-downstream phase correlation to \(98\%\) in typical conditions, and 
places a theoretical limit of \(0.17^\circ\) on the measurable corrected 
downstream phase jitter. 
The dominant beam source of uncorrelated downstream phase jitter 
arises from energy jitter that is transformed into phase jitter in the 
beamlines between \(M_1\) and \(M_3\). 

\begin{figure}
	\includegraphics[width=\columnwidth]{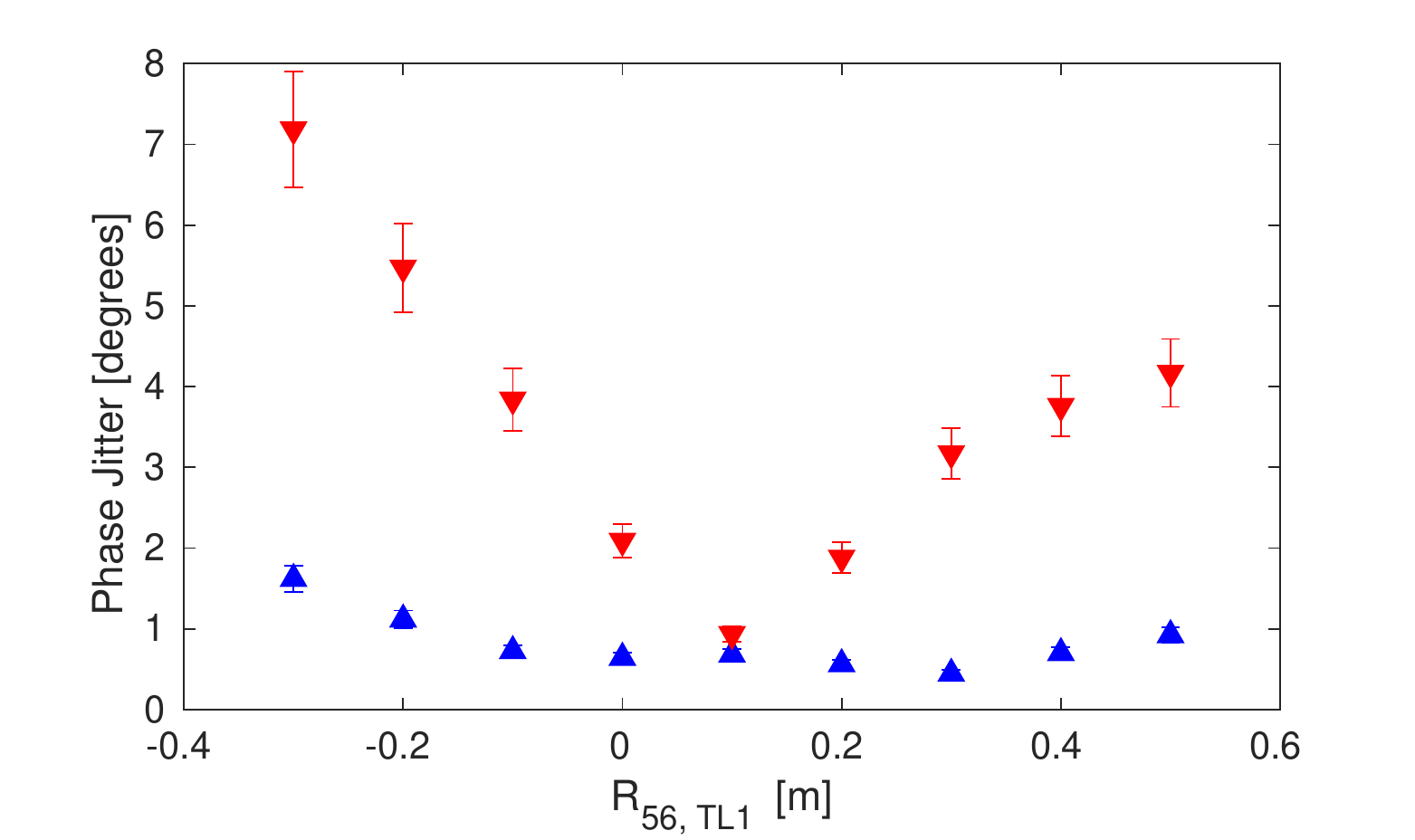}
	\caption{\label{fig:r56Scan}Measured downstream (red) and upstream (blue) 
	phase jitter vs. TL1 \(R_{56}\) value. Error bars show the statistical 
	standard  error on the measured jitter values.
		}
\end{figure}

To first order the phase dependence on energy can be described via the beam 
transfer matrix coefficient 
\(R_{56}\): \(\phi_3 = \phi_1 + \alpha R_{56}(\Delta p / p)\)
, where \(\Delta p / p\) is the particle's relative energy error, \(\phi_1\) 
and \(\phi_3\) are the phases measured at \(M_1\) and \(M_3\) respectively, 
and the constant \(\alpha = 14400\)~\(^\circ/\)m converts the units of 
\(R_{56}\) from metres to degrees at 12~GHz (\(360^\circ\) per 0.025~m).

The optimal condition is \(R_{56}\) = 0.
This was achieved by tuning the \(R_{56}\) value in the `TL1' transfer line 
(Fig.~\ref{fig:pffLayout}) so as to compensate for non-zero \(R_{56}\) in the 
other beamline sections. With \(R_{56, \mathrm{TL1}}=10\)~cm the 
downstream phase jitter is reduced to the same level as the upstream jitter 
(Fig.~\ref{fig:r56Scan}). 
However, a large \(R_{566}\) coefficient (second-order phase dependence on 
energy) remained uncorrected.
As a result, drifts in beam energy lead to a degradation in upstream-downstream 
phase correlation even after optimising the \(R_{56}\) term.
Drifts in the CTF3 RF system, and the resulting changes in beam energy, 
therefore made it difficult to maintain maximal upstream-downstream phase 
correlation for timescales longer than 10 minutes. Optics for a future CLIC PFF 
system must zero both \(R_{56}\) and the higher order energy dependences.

\begin{figure}
\includegraphics[width=\columnwidth]{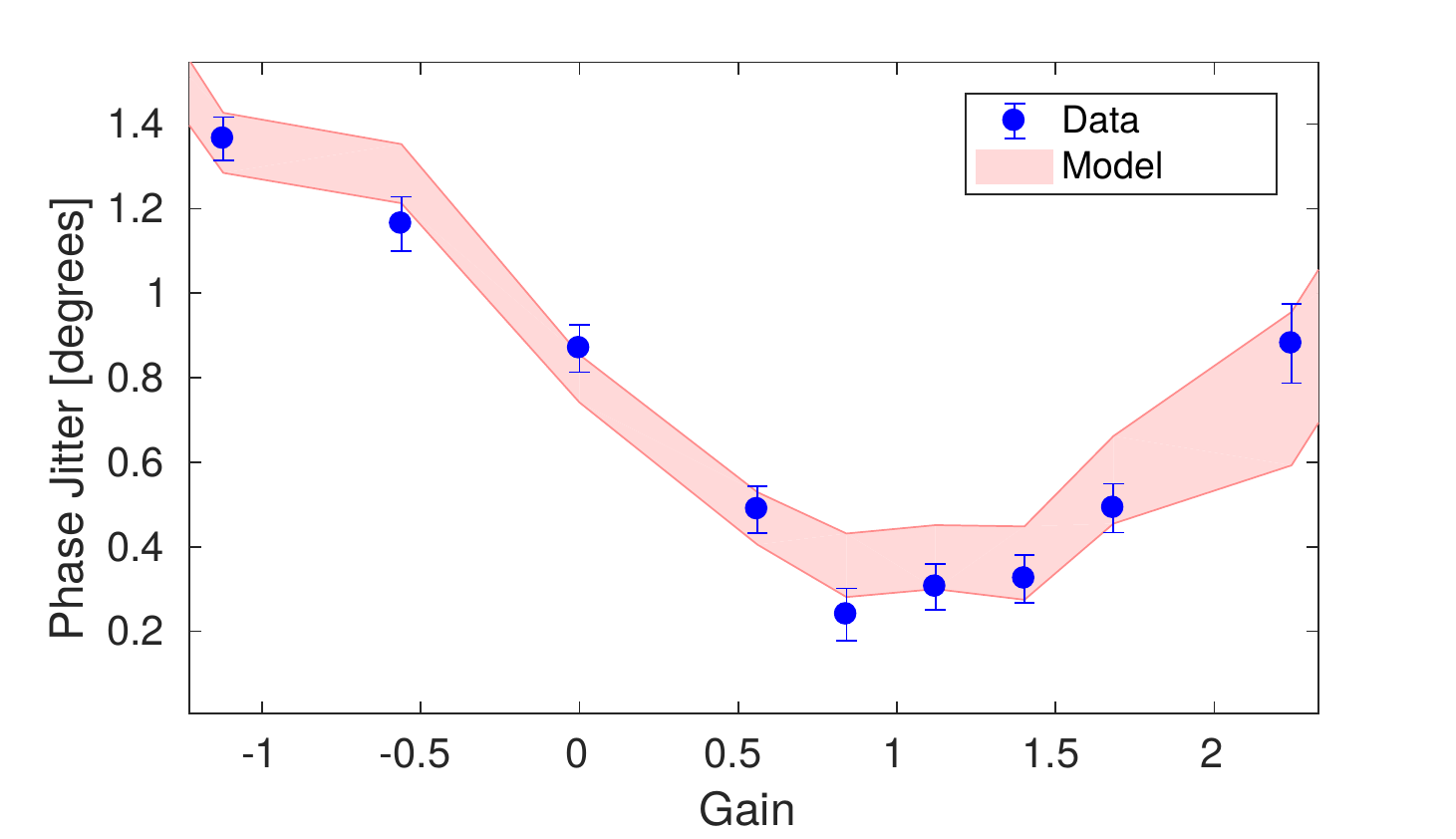}
\caption{\label{fig:gScan}Measured corrected beam  phase jitter at \(M_3\) 
vs. PFF gain (points). Error bars show the statistical standard error on the 
measured jitter values. The expected 
performance is shown by the red shaded region (see text).}
\end{figure}

The PFF system acts to remove the \(M_1\) phase, multiplied by a `gain' 
factor, from the phase at \(M_3\). If the phases at \(M_3\) and 
\(M_1\) are fully correlated, and the jitters are identical, the optimal 
system gain is unity.
In practice the gain is chosen to achieve optimal 
performance for real beam conditions. A representative gain scan is shown 
in~Fig.~\ref{fig:gScan}. The optimal gain is typically in the range 
0.9--1.3. Also shown in Fig.~\ref{fig:gScan} is a prediction of 
the corrected phase jitter at \(M_3\), using a simple model including the 
initial beam phase jitters at \(M_1\) and 
\(M_3\), the upstream-downstream phase correlation, and the gain 
\cite{RobertsThesis}. The model reproduces the data.

\begin{figure}
	\begin{subfigure}{\columnwidth}
		\includegraphics[width=\textwidth]{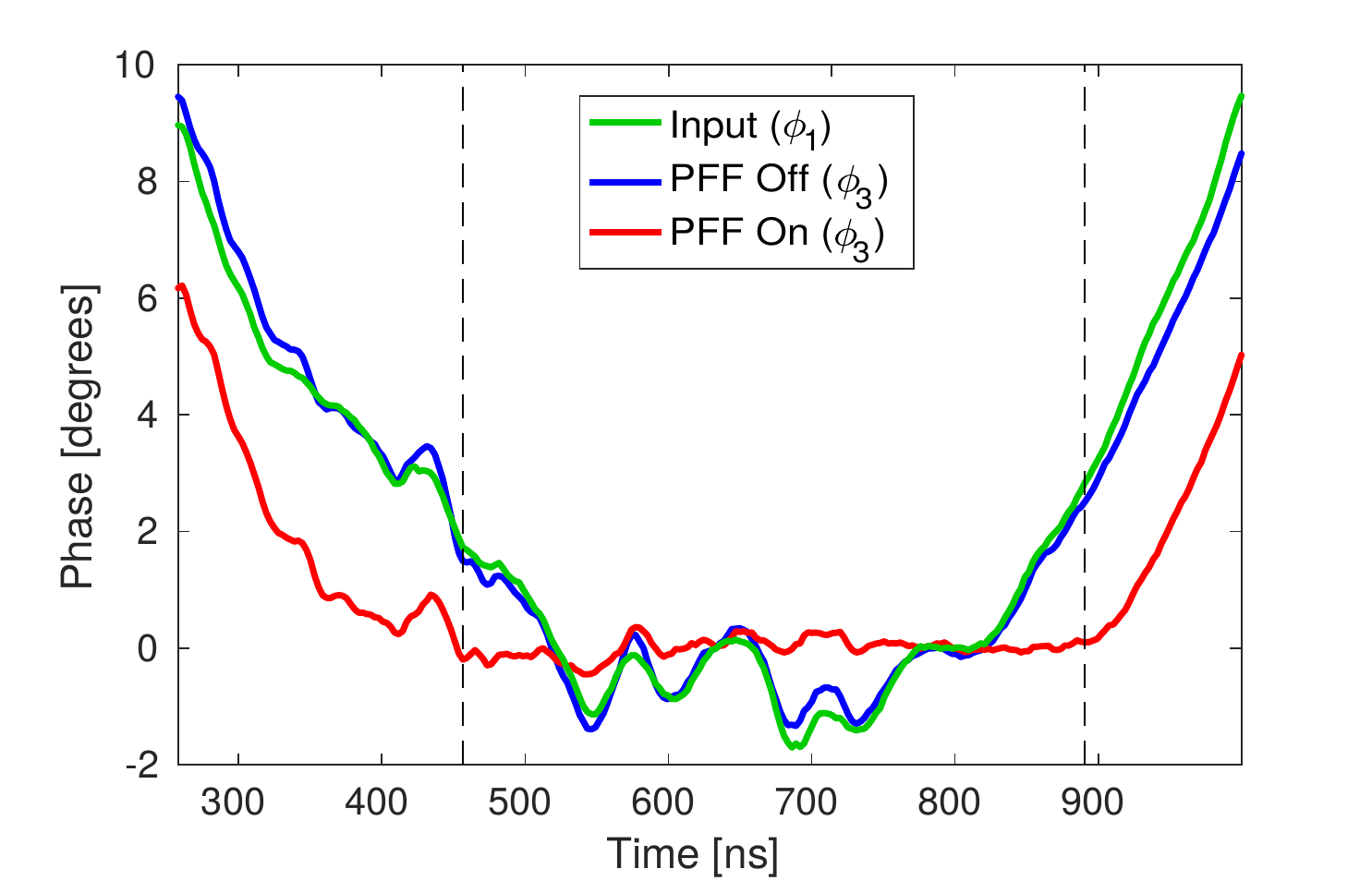}
		\caption{}
		\label{f:shapeAll}
	\end{subfigure}
	
	\begin{subfigure}{\columnwidth}
		\includegraphics[width=\textwidth]{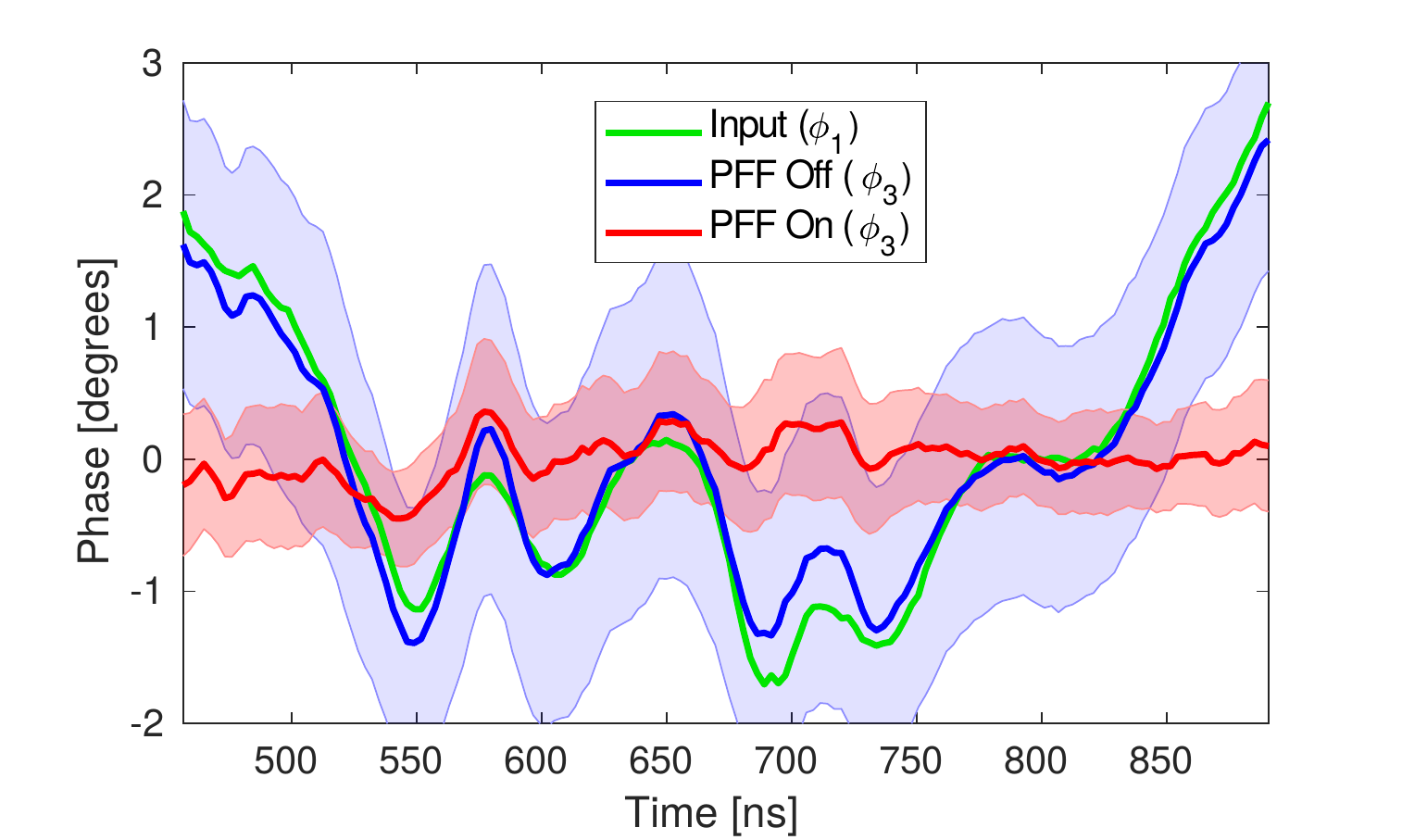}
		\caption{}
		\label{f:shapeZoom}
	\end{subfigure}	
	\caption[LoF entry]{\label{fig:shape}Correction of the pulse shape with the 
	PFF system. Shown are: the incoming phase (\(\phi_1\)) measured in \(M_1\)  
		(green), and the downstream phase (\(\phi_3\)) measured in \(M_3\) with 
		PFF off (blue) and PFF on (red). Each trace is the average 
		over a 30 minute dataset. 
		
		(a) 		
		The whole beam pulse. Vertical dashed 
		lines mark the time interval corresponding to the PFF 
		dynamic range.		
		(b) 		
		The same data zoomed in to the central portion of the pulse. 
		Shaded areas represent the phase jitter at each sample	point.		
	}
\end{figure}

The PFF system simultaneously corrects pulse-to-pulse phase jitter and phase 
variations within the 1.2~\(\mu s\) beam pulse at CTF3. 
Fig.~\ref{fig:shape} shows the effect of the PFF system on the intra-pulse 
phase variations. The PFF system was operated in interleaved mode, with 
the correction applied to alternating pulses only. This allows 
the initial (`PFF Off') and corrected (`PFF On') downstream phase at \(M_3\)
to be measured at the same time. The \(M_1\) (PFF input) phase 
is also shown for comparison. 

It is an operational feature at CTF3 that there is a roughly parabolic phase 
sag of \(40^\circ\) along the pulse, resulting from the upstream RF pulse 
compression 
scheme~\cite{CLICCDR}. Hence approximately a 440~ns portion of the pulse is 
within the \(\pm 6^\circ\) dynamic range of the PFF system, and can be 
corrected to zero nominal phase. 
This time duration for the full correction exceeds the CLIC drive-beam pulse 
length of 240ns and in any case the CLIC design avoids such 
a large phase sag~\cite{CLICCDR}. 
Vertical dashed lines in Fig.~\ref{fig:shape} mark the 440~ns portion of 
the pulse where full correction is possible.


Within the range the PFF system flattens the phase, and almost all variations 
are removed. 
The average intra-pulse phase variation (rms) over the dataset is reduced from 
\(0.960\pm0.003^\circ\) (PFF off), to \(0.285\pm0.004^\circ\) (PFF on).

\begin{figure}
	\includegraphics[width=\columnwidth]{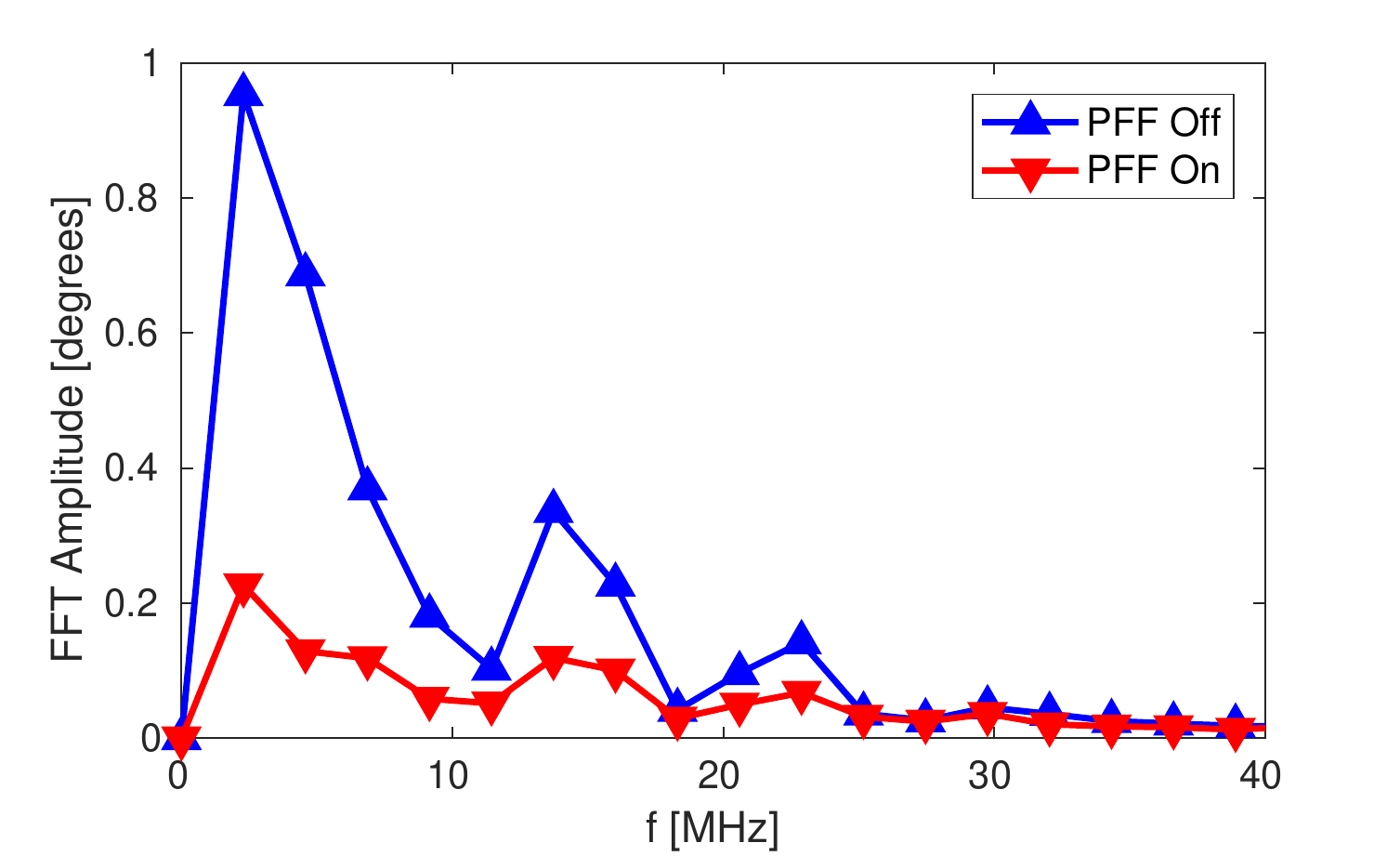}
	\caption{\label{fig:fft}Amplitude of phase errors vs. frequency (\(f\)) in 
	bins of 2.3~MHz with the PFF system off (blue) and on (red), across a 30 
	minute 
		dataset.}
\end{figure}

In order to meet CLIC requirements (Table~\ref{tab:pffspecs}) the PFF 
correction bandwidth should be at least 17.5 MHz. 
A Fourier-Transform (FFT) method was used to characterise the PFF on/off 
datasets. The FFT amplitude is shown vs. frequency in 
Fig.~\ref{fig:fft}. It can be seen that phase errors are corrected by up to a 
factor of 5 for frequencies up to 23~MHz, above which 
they are smaller than the monitor resolution and not measurable. This 
is consistent with an expected system bandwidth of around 30~MHz, and exceeds 
the CLIC requirement.

\begin{figure}
	\includegraphics[width=\columnwidth]{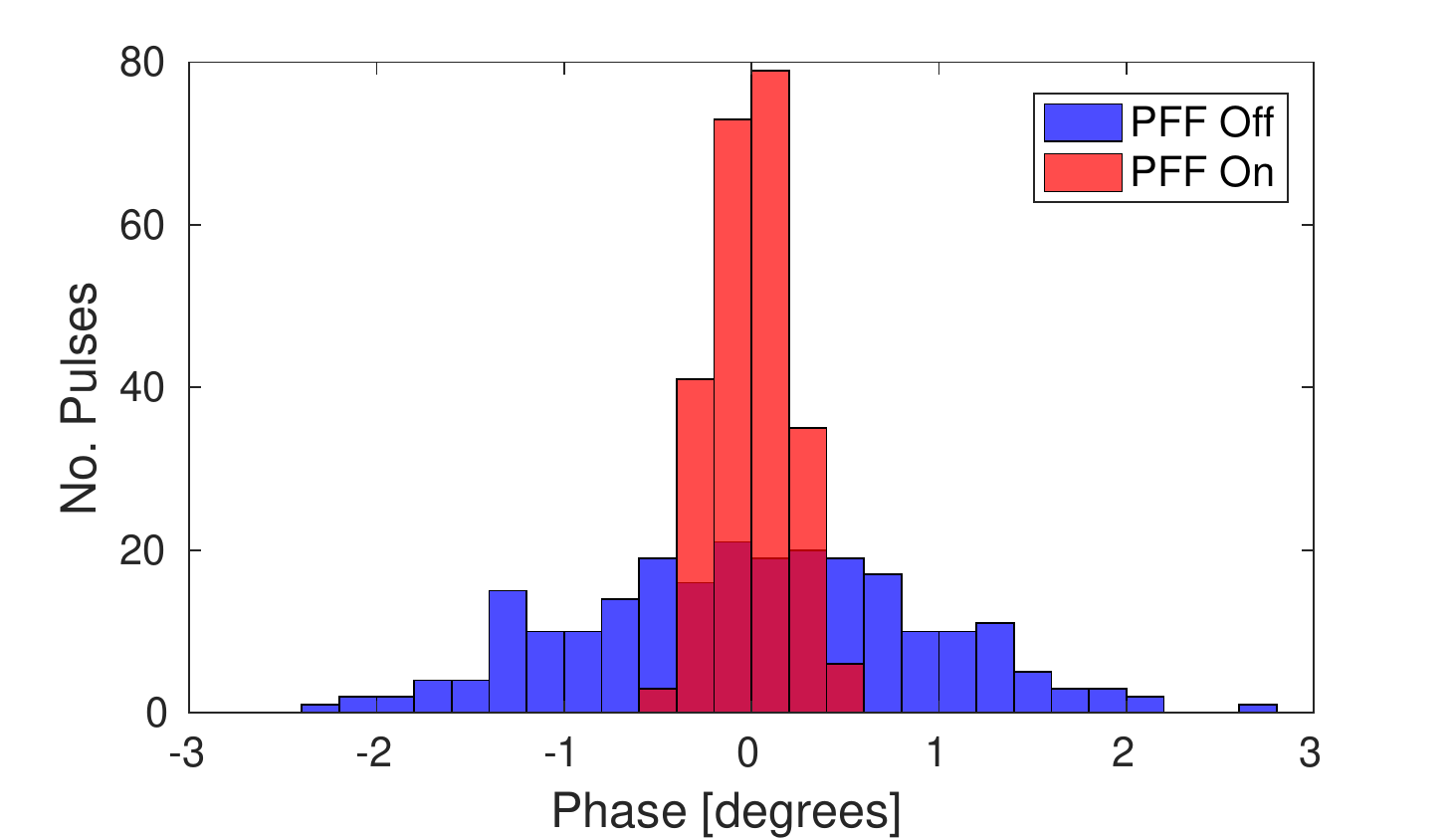}
	\caption{\label{fig:meanJit}Distribution of the mean downstream phase with 
		the 
		PFF system off (blue) and on (red).}
\end{figure}

The effect of the PFF system on the pulse-to-pulse jitter, i.e. the jitter on 
the mean phase of each beam pulse, is shown in Fig~\ref{fig:meanJit} for a 
dataset of around ten minutes duration.
The pulse-to-pulse phase jitter is reduced from  \(0.92\pm0.04^\circ\) to 
\(0.20\pm0.01^\circ\), meeting CLIC-level phase stability. 
The system acts to remove all correlations between the upstream and 
downstream phase, reducing an initial correlation of \(96\pm2\%\) to 
\(0\pm7\%\) for this dataset.
Given the incoming upstream phase jitter and 
measured upstream-downstream correlation, the performance is consistent with 
the theoretically predicted correction of \(0.26\pm0.06^\circ\).

\begin{figure}
	\includegraphics[width=\columnwidth]{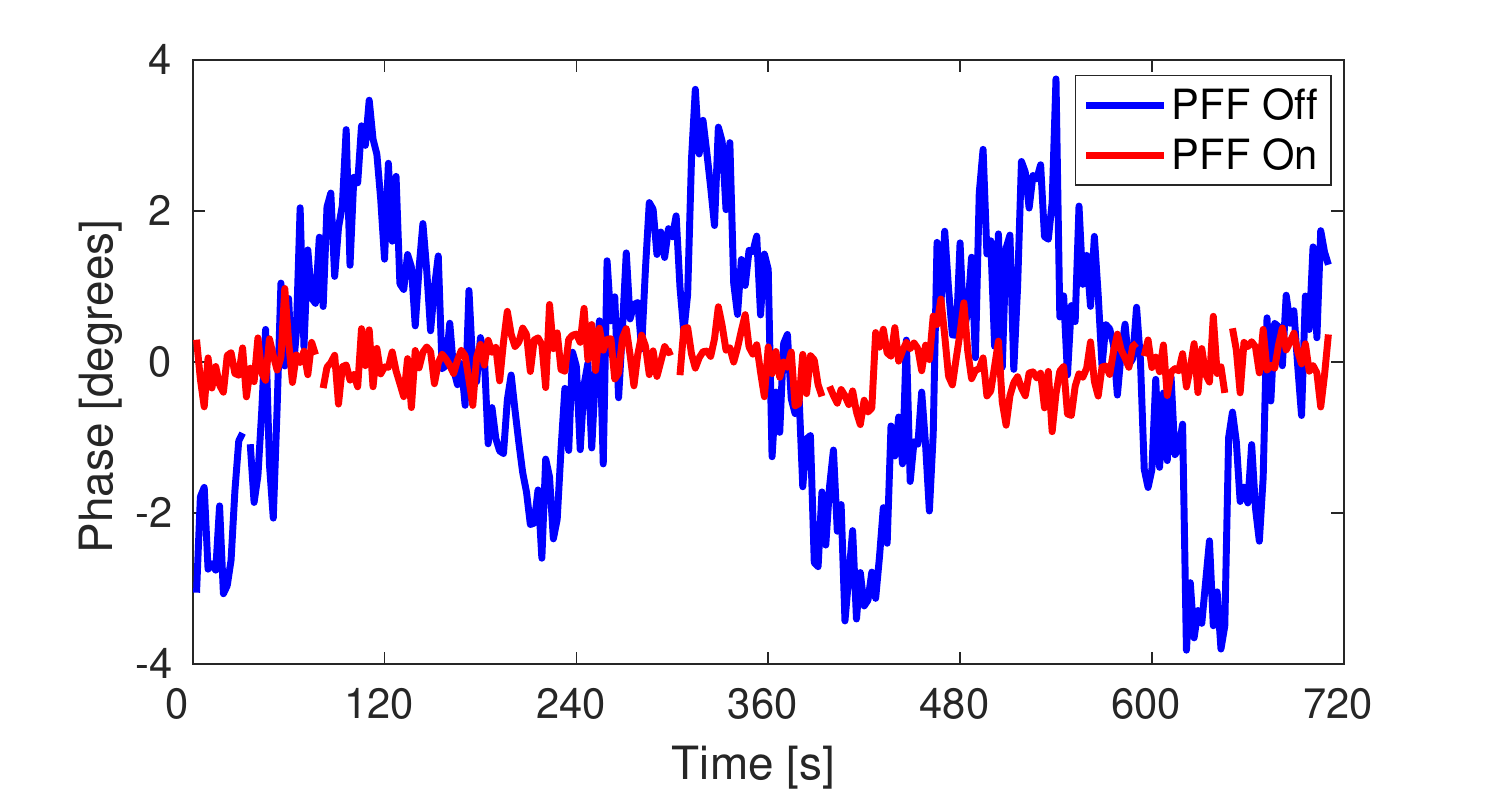}
	\caption{\label{fig:wiggle}Mean downstream phase vs. time with the PFF 
	system off (blue) and on (red) subject to large additional phase variations 
	added to the incoming phase (see text).}
\end{figure}

The system was further tested by varying the incoming mean 
beam phase systematically by around \(\pm 3^\circ\) (Fig.~\ref{fig:wiggle}). 
Variations of this magnitude 
are comparable to the expected conditions in the CLIC design 
(Table~\ref{tab:pffspecs}). This is illustrated in Fig.~\ref{fig:wiggle}. The 
system removed the induced phase variations and achieved more than a factor-5 
reduction in the downstream phase jitter, 
correcting~from~\(1.71\pm0.07^\circ\)~to~\(0.32\pm0.01^\circ\). 

In summary, we have built, deployed and tested a prototype drive-beam phase 
feedforward system for CLIC.   The system incorporates purpose-built 
high-resolution phase 
monitors, an advanced signal-processor and feedforward controller, low-latency, 
high-power, high-bandwidth amplifiers, and electromagnetic stripline
kickers. The phase-monitor resolution was measured to be 
\(0.12^\circ\simeq\)~30~fs.  The overall system latency, including the hardware 
and signal transit times, was measured to be approximately 350~ns, which is 
less than 
the beam time of flight between the input phase monitor and the correction 
chicane.
The system was used to 
stabilise the pulse-to-pulse phase jitter to \(0.20\pm0.01^\circ\simeq\)~50 fs, 
and to simultaneously correct intra-pulse phase variations at frequencies up to 
23~MHz.

Our demonstration of a beam-based arrival-time stabilisation system with a 
performance at the 50 fs level has potential application at other beamlines 
where a high degree of beam arrival stability is required.  For example, 
`pump-probe' experiments at FELs require laser/electron synchronisation ideally 
to the few femtosecond level, see e.g. \cite{Savelyev2017}. The current 
state-of-the-art in synchronisation at FELs is approximately 30~fs, using 
all-optical 
techniques \cite{FLASHnature}. Our results are limited by the beam arrival-time 
monitor resolution of approximately 30~fs. With higher precision monitors (e.g. 
\cite{flashPRL}) 10~fs stabilisation could be achieved with our technique. A 
key feature of our system is that it incorporates a beam turnaround, which 
provides sufficient beam delay to allow a feed-forward correction to be derived 
and applied with zero effective latency. FEL designs based on energy-recovery 
linacs (see e.g. \cite{Sekutowicz2005,Kwang2008,Jackson2016}) intrinsically 
incorporate a beam turnaround section that would enable the deployment of a 
high-performance system based on our technique.

\begin{acknowledgments}
	We acknowledge Alessandro Zolla and Giancarlo Sensolini (INFN 
	Frascati) for the mechanical design of the phase monitors and 
	kickers, and Alexandra Andersson, Luca Timeo and Stephane Rey (CERN) for 
	their support on the phase monitor electronics. We thank the 
	operations team of CTF3 for their outstanding support. This work was  
	supported by the the UK Science and Technology Facilities 
	Council, and by the European Commission under the FP7 Research 
	Infrastructures project Eu-CARD, grant agreement no. 227579.
\end{acknowledgments}

\end{document}